\documentclass[doublecol]{epl2} 
\usepackage{textcomp}

\title{Field-induced ferromagnetism in one-dimensional tight-binding lattices}
\shorttitle{Field-induced ferromagnetism in 1D} 

\author{Giuseppe Della Valle\footnote{giuseppe.dellavalle@polimi.it} \and Stefano Longhi}
\shortauthor{G. Della Valle \etal}

\institute{                    
Dipartimento di Fisica, Politecnico di Milano and \\Istituto di Fotonica e Nanotecnologie del Consiglio Nazionale delle Ricerche\\ Piazza L. da Vinci 32, I-20133 Milano, Italy\\
}

\abstract{
We theoretically show the possibility to induce magnetic ordering in non-magnetic one-dimensional systems of strongly interacting electrons hopping on a tight-binding lattice. Our analysis is provided within the framework of the $t_1$-$t_2$ Hubbard Model, assuming non-zero second neighbor hopping rate. It is shown that a high-frequency electric field can be exploited to induce artificial ferromagnetism and eventually control anti-ferromagnetic/ferromagnetic phase transition. Our analysis is validated by numerical simulations in a low-density system of 2-particles on a lattice with 11 sites.
}

\begin{document}

\maketitle

\section{Introduction}
The appearance of ferromagnetism in strongly-interacting quantum systems is the subject of a long-lasting research in solid-state physics. Since the early sixties, many efforts have been devoted to prove the existence of large-spin and eventually saturated-spin ground state in Hubbard-type models \cite{Hubbard_63,Kanamori_63}. Nagaoka ferromagnetism has been demonstrated for one hole in an half-filled band in the limit of infinite electron interaction \cite{Nagaoka_66}. Lieb ferromagnetism for half-filled bipartite lattices \cite{Lieb_89} and Mielke-Tasaki ferromagnetism in flat band systems \cite{Tasaki_92,Mielke_93,Mielke-Tasaki_93} are other important cases where the appearance of ferromagnetism has been rigorously established. Another approach to the theoretical description of ferromagnetism within the Hubbard model considers both nearest-neighbor hopping rate $t_1$ and next-nearest-neighbor hopping rate $t_2$ (the so-called $t_1$-$t_2$ Hubbard model). This led to the demonstration of the M\"uller-Hartmann ferromagnetism in the limit of infinite interaction $U$ and low electron density $n$ \cite{Muller-Hartmann_95}. A generalization of the latter scenario to a finite interaction energy $U$ and for arbitrary particle density $n$ has been reported by Pieri and coworkers in Ref.~\cite{Pieri_96}. In that work, extended numerical simulations of the one-dimensional $t_1$-$t_2$ Hubbard model indicate the existence of fully polarized ferromagnetic states in an extended region of the $U-n$ plane, provided that $t_1 t_2 <0$ and the ratio $-t_2/t_1$ is sufficiently large \cite{Pieri_96}. More recent applications of the one-dimensional $t_1$-$t_2$ Hubbard model to ferromagnetism have led to the demonstration that the order of ferromagnetic transition is determined by the quantum-liquid phase of the system \cite{Daul_00}. Ferromagnetic ordering has been demonstrated in Wigner lattices, where long-range Coulomb repulsion dominates over kinetic energy of electrons \cite{Daghofer_08}. Also, the onset of $t_1$-$t_2$ Hubbard ferromagnetism has been predicted and thoroughly investigated in two-dimensional lattices \cite{Taniguchi_PRB_05,Pandey_PRB_07}.

Though the $t_1$-$t_2$ route to ferromagnetism has been widely explored, in all previous studies a static (i.e. time-independent) Hubbard Hamiltonian has been considered and no attempts to investigate the effects of an external driving field on the magnetic ordering within $t_1$-$t_2$ Hubbard model have been so far reported.

On the other hand, time-dependent Hubbard models, describing the effects of external driving fields or parameter modulations, can provide a fertile ground to control the properties of correlated-particles systems. Examples include field-controlled superfluid to Mott-insulator phase transition~\cite{Longhi_PRB_08,Zenesini_PRL_08}, dynamic unbinding transitions in a periodically-driven fermionic Mott-insulator at half-filling~\cite{Hassler_PRL_10}, switching of the interaction from repulsive to attractive~\cite{Tsuji_PRL_11}, and control of correlated tunneling and superexchange spin interactions by ac fields~\cite{Chen_PRL_11}.

In this paper we consider a $t_1$-$t_2$ Hubbard model describing the hopping motion of $N$ interacting electrons on a one-dimensional lattice driven by an external ac electric field. The main result of our analysis is that, in the high frequency regime, the time-dependent Hubbard model results in an effective static Hubbard model with renormalized $t'_1$-$t'_2$ hopping rates. The capability to control the magnitude and relative sign of $t'_1$ and $t'_2$ by the external field in a broad range of values shows the possibility to induce magnetic ordering transitions of the ground state of the system. The predictions of the asymptotic analysis are confirmed by direct numerical simulations of the time-periodic Hubbard model in case of $N=2$ interacting electrons.

\section{The driven $\bf t_1$-$\bf t_2$ Hubbard model}

In the presence of an external driving electric field $E(t)$, the $t_1$-$t_2$ Hubbard Hamiltonian of a one-dimensional system of interacting electrons reads as follows:
\begin{equation}
\hat{H} = \hat{H}_{hop1}+\hat{H}_{hop2}+\hat{H}_{int}+\hat{H}_{drive}
\end{equation}
\noindent where
\begin{eqnarray}
\hat{H}_{hop1}&=& - \hbar t_1 \sum_{j=1}^{L-1} \sum_{\sigma=\uparrow,\downarrow}\left( \hat{a}^{\dag}_{j,\sigma}\hat{a}_{j+1,\sigma}+\hat{a}^{\dag}_{j+1,\sigma}\hat{a}_{j,\sigma}  \right)\\
\hat{H}_{hop2}&=& -  \hbar t_2 \sum_{j=1}^{L-2} \sum_{\sigma=\uparrow,\downarrow}\left( \hat{a}^{\dag}_{j,\sigma}\hat{a}_{j+2,\sigma}+\hat{a}^{\dag}_{j+2,\sigma}\hat{a}_{j,\sigma}  \right)\\
\hat{H}_{int}&=&  U \sum_{j=1}^{L} \prod_{\sigma=\uparrow,\downarrow}\hat{n}_{j,\sigma}\\
\hat{H}_{drive}&=& edE(t) \sum_{j=1}^{L} \sum_{\sigma=\uparrow,\downarrow} j \hat{n}_{j,\sigma}.
\end{eqnarray}
In previous equations, $L$ is the number of lattice sites, $d$ is the lattice period, $U$ is the on-site Coulomb repulsion energy, $\hat{a}^{\dag}_{j,\sigma}$ is the fermionic creation operator that creates one electron at site $j$ with spin $\sigma$ ($j = 1,2,...,L; \sigma =\uparrow,\downarrow$), and $\hat{n}_{j,\sigma}=\hat{a}^{\dag}_{j,\sigma}\hat{a}_{j,\sigma}$ is the spin $\sigma$ particle number operator at lattice site $j$.

The state vector of the system $|\psi(t) \rangle$ in Fock space representation can be written as:
\begin{equation}
| \psi(t) \rangle= \sum_{{\rm \bf n,m}}f({\rm \bf n,m},t) |{\rm \bf n,m} \rangle,
\end{equation}
\noindent where $f({\rm \bf n,m},t)$ is the complex amplitude for the $|{\rm \bf n,m} \rangle = | n_1,n_2,...n_L,m_1,m_2,...m_L \rangle$ Fock basis element, representing a state with $n_j$ electrons occupying the $j$ lattice site with spin $\downarrow$ and $m_j$ electrons occupying the $j$ lattice site with spin $\uparrow$, with $n_j,m_j$ taking only the two values 0 and 1, owing to the anti-commutation rules of the Fermi operators. Given above decomposition of the state vector, standard projection technique provides the following evolution equation for the amplitudes $f({\rm \bf n,m},t)$:
\begin{equation}
{\rm i}\hbar \frac{df({\rm \bf n,m},t)}{dt}= \sum_{{\rm \bf s,q}} \langle {\rm \bf n,m}|\hat{H}|{\rm \bf s,q} \rangle f({\rm \bf s,q},t).
\end{equation}

\section{Renormalized $\bf t'_1$-$\bf t'_2$ Hubbard model and field-induced magnetic ordering transition}

The dynamics of the system under a high-frequency driving field can be at best captured after the substitution:
\begin{equation}
f({\rm \bf n,m},t) = g({\rm \bf n,m},t) \exp \left[-{\rm i}\Phi(t) \sum_{j=1}^{L} j(n_j+m_j)\right],
\end{equation}
\noindent where
\begin{equation}
\Phi(t)=\frac{ed}{\hbar}\int_0^t dt' E(t').
\end{equation}
This way, the amplitude probabilities $g({\rm \bf n,m},t)$ satisfy the following coupled equations:
\begin{eqnarray}
{\rm i} \frac{dg({\rm \bf n,m},t)}{dt}= \frac{1}{\hbar} \sum_{{\rm \bf s,q}} \langle {\rm \bf n,m}|\hat{H}_{hop1}+\hat{H}_{hop2}|{\rm \bf s,q} \rangle \\ \nonumber 
\times \exp \left[ {{\rm i} \rho({\rm \bf n,m,s,q})\Phi(t)}\right] g({\rm \bf s,q},t)\\ \nonumber 
+ \frac{1}{\hbar} \sum_{{\rm \bf s,q}} \langle {\rm \bf n,m}|\hat{H}_{int}|{\rm \bf s,q} \rangle g({\rm \bf s,q},t).
\end{eqnarray}
In Eq.~(10) we have set
\begin{equation}
\rho({\rm \bf n,m,s,q})= \sum_{j=1}^L j(n_j-s_j+m_j-q_j).
\end{equation}
Note that since the effect of the tunneling Hamiltonians $\hat{H}_{hop1}$ and $\hat{H}_{hop2}$ is to shift one electron from lattice site $j$ to lattice site $j\pm1$ or $j\pm2$ respectively, the nonvanishing matrix elements entering in Eq.~(10) correspond to $\rho({\rm \bf n,m,s,q})=\pm1$ for $\hat{H}_{hop1}$ and to $\rho({\rm \bf n,m,s,q})=\pm2$ for $\hat{H}_{hop2}$.
If we now assume that the external driving field is periodic with period $T = 2\pi / \omega$, in the high-frequency limit $\omega \gg t_1,t_2, U$, the
rapidly oscillating exponential terms on the right-hand side of Eq.~(10) can be replaced by their time average over one oscillation cycle of the ac field \cite{Longhi_PRB_08}, resulting in an effective renormalization of the hopping amplitudes \cite{Eckardt_05}. In particular, for a sinusoidal driving field $E(t) = E_0 \sin(\omega t)$, the averaging procedure yields for the renormalized hopping rates $t'_1,t'_2$ the following expressions:
\begin{eqnarray}
t'_1 &=& t_1\langle \exp[\pm {\rm i} \Phi(t)] \rangle = t_1J_0(\Gamma)\\
t'_2 &=& t_2\langle \exp[\pm {\rm i} 2\Phi(t)] \rangle = t_2J_0(2\Gamma)
\end{eqnarray}
\noindent where $\langle...\rangle = 1/T \int_0^T dt...$, $\Gamma = edE_0 / (\hbar \omega)$, and $J_0$ is the Bessel function of first kind and zero order.
Note that the ratio $t'_2/t'_1=r t_2/t_1$ with $r=J_0(2\Gamma)/J_0(\Gamma)$, can be made negative and arbitrarily large in modulus provided that $z_0/2<\Gamma<z_0$, being $z_0\simeq 2.405$ the first zero of $J_0$, (Fig.~1). Hence the effect of the driving field is basically to renormalize the original $t_1$ and $t_2$ hopping rates, with a renormalization ratio $r$ that can be tuned over a wide range of values.
\begin{figure}[ht]
\begin{center}
\onefigure[width=8cm]{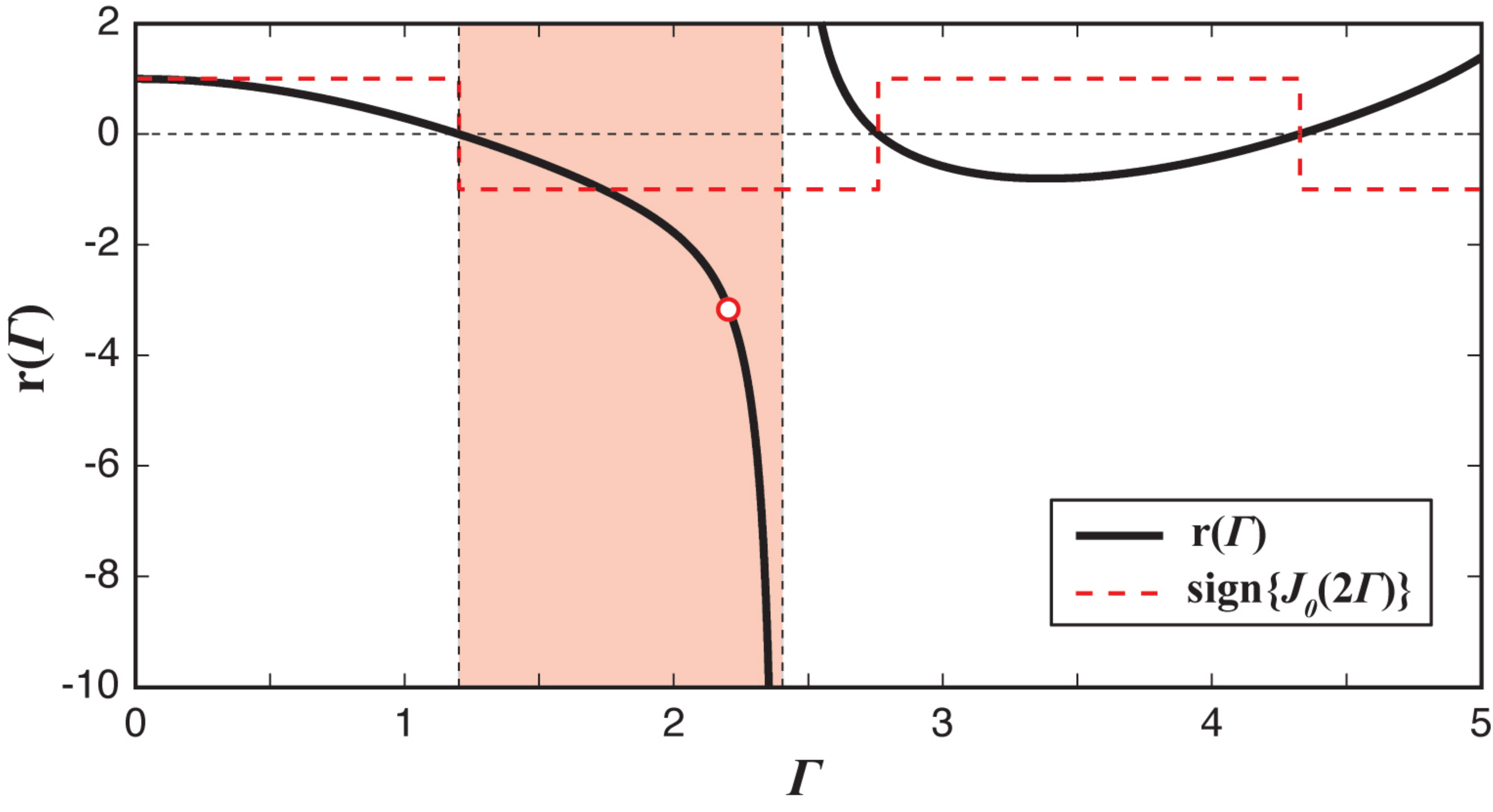}
\caption{The renormalization ratio $r$ between second and first neighbor tunneling rates as a function of $\Gamma$. The red shaded area corresponds to $z_0/2<\Gamma<z_0$. Red circle represents the value $\Gamma = 2.2$ used in the simulations of Fig.~2.}
\end{center}
\label{Fig1}
\end{figure}
Our analysis is valid for an arbitrary density of electrons $n=N/L$, and thus the results reported in Ref. \cite{Pieri_96} can be readily applied to the effective $t'_1$-$t'_2$ Hubbard Hamiltonian. In particular, it is known that for any given density of electrons and for any given interaction energy $U$, there exists a threshold value for the ratio $-t'_2/t'_1$ above which the ground state of the system becomes ferromagnetic~\cite{Pieri_96}.
By proper control of the ratio $E_0/\omega$ we can thus actively turn the ground state of the system into a ferromagnetic state, regardless the value of the hopping rates $t_1,t_2$ and of the interaction energy $U$ of the original (static) Hubbard Hamiltonian.

\section{Numerical results and discussion}

To validate the previous analysis we consider the simplest case of $N=2$ electrons with opposite spins, for which numerical simulations can be easily performed for a reasonably large number of lattice sites. In this case, Fock space representation of the state vector of the system reads as follows:
\begin{equation}
| \psi(t) \rangle= \sum_{n,m} c_{n,m}(t) \hat{a}^{\dag}_{n,\downarrow} \hat{a}^{\dag}_{m,\uparrow}|{\rm 0} \rangle,
\end{equation}
\noindent where $c_{n,m}(t)$ is the complex amplitude for the basis state corresponding to one electron with spin $\downarrow$ at lattice site $n$ and one electron with spin $\uparrow$ at site $m$. By inserting the ansatz of Eq.~(14) into previous Eqs.~(1)-(5), we end up with the following evolution equations for the amplitude probabilities $c_{n,m}$:
\begin{eqnarray}
{\rm i} \frac{dc_{n,m}}{dt} &=& - t_1 \left( c_{n-1,m}+c_{n+1,m}+c_{n,m-1}+c_{n,m+1} \right) \nonumber \\
&-& t_2 \left( c_{n-2,m}+c_{n+2,m}+c_{n,m-2}+c_{n,m+2} \right) \nonumber \\
&+& \left[ \frac{U}{\hbar}\delta_{n,m}+\Gamma \omega(n+m) \sin(\omega t) \right] c_{n,m}.
\end{eqnarray}
\begin{figure}[ht]
\begin{center}
\onefigure[width=8cm]{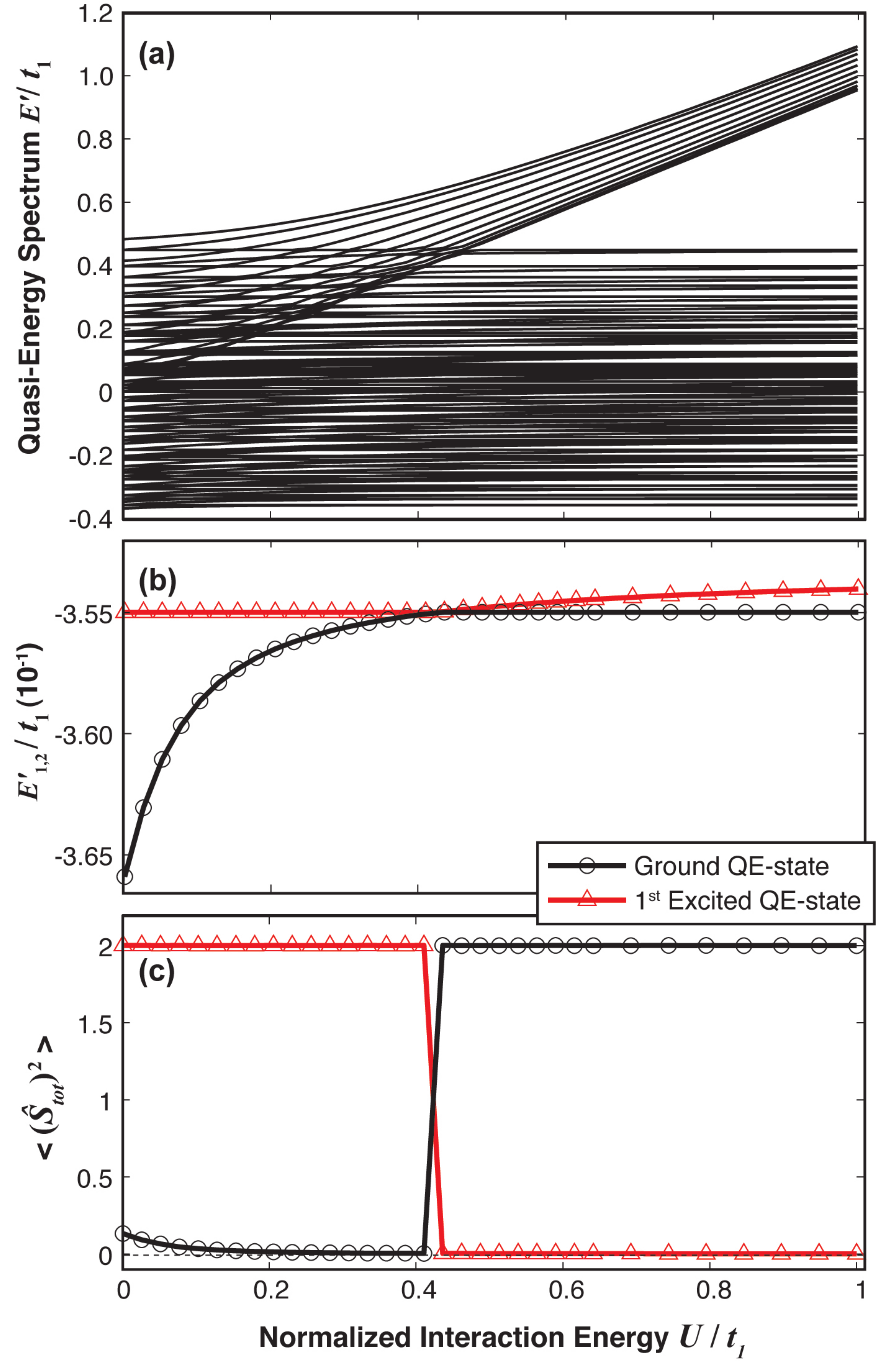}
\caption{(a) Quasi-energy spectrum of the driven $t_1$-$t_2$ lattice as a function of interaction energy. (b) Zoom-in of the first two quasi-energies. (c) Expectation value of the total spin operator $\hat{S}_{tot}^2$ for the first two quasi-energy (QE) states of the system.}
\end{center}
\label{Fig2}
\end{figure}
The behaviour of the two-electron system under high frequency driving with a strong electric field can be understood by inspecting the quasi-energies and quasi-energy states of the system described by Eq.~(15), computed according to standard Floquet theory~\cite{Grifoni_PR_98}. 
\begin{figure}[ht]
\begin{center}
\onefigure[width=8cm]{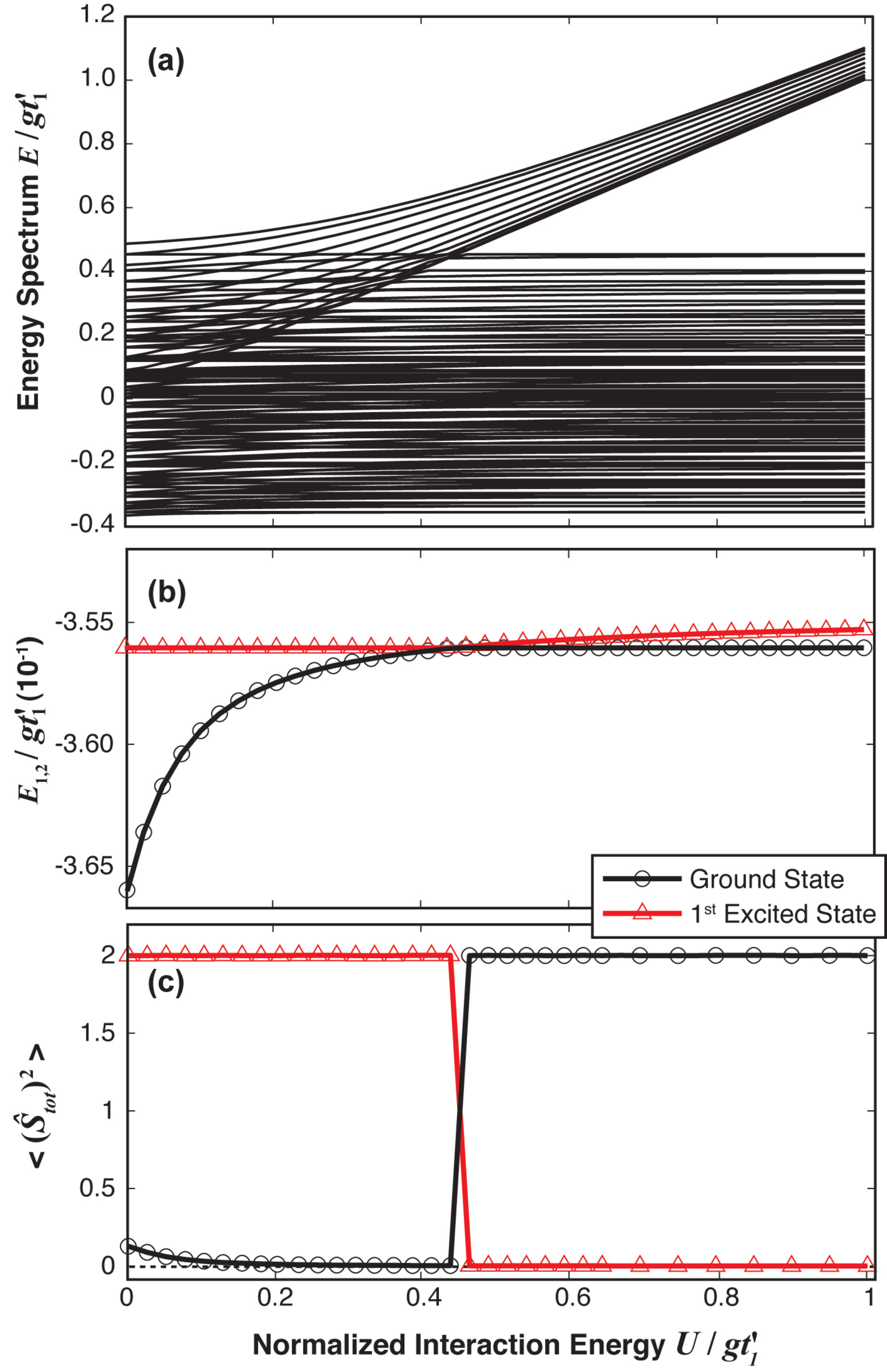}
\caption{(a) Energy spectrum of the equivalent $t'_1$-$t'_2$ static lattice as a function of interaction energy. (b) Zoom-in of the first two energy levels. (c) Expectation value of the total spin operator $\hat{S}_{tot}^2$ for the first two eigenstates of the system. Energy values are normalized to $g t'_1 = t_1$ being $g = 1/J_0(\Gamma)$, for a better comparison with the results of Fig.~2.}
\end{center}
\label{Fig3}
\end{figure}
As an example, in Fig.~2(a) we show the computed quasi-energy spectrum $E'$ for parameter values $t_2 = 0.05 t_1$, $\omega = 10 t_1$ and $\Gamma = 2.2$ (see red circle in Fig.~1), with $U$ ranging between $0$ and $t_1$. Within the $t_1$-$t_2$ Hubbard model, an increase of $U$ above a critical value is known to be responsible for the onset of ferromagnetism~\cite{Pieri_96,Tasaki_98}. The quasi-energy spectrum shown in Fig.~2(a) clearly indicates a typical Mott-Hubbard band-gap formation at a certain value of the interaction energy, with a bounce of higher levels emerging from a broader spectrum of lower lying levels. These higher levels correspond to bound-particle states (doublons). The onset of magnetic ordering in the system can be investigated by computing the expectation value of the total spin operator $\hat{S}^2_{tot}$, whose eigenvalues are given by $S_{tot}(S_{tot}+1)$, with $S_{tot}$ the total spin of the system, that ranges from 0 (anti-ferromagnetic state) to a maximum value $S_{max}$ (saturated ferromagnetic state) that depends on the number of particles, namely $S_{max} = N/2$ or $S_{max} = L-N/2$ respectively below or above particle density $n = 1$, that is half-filling \cite{Tasaki_98}. 
A zoom-in of the spectrum showing the first two quasi-energies is reported in Fig.~2(b), whereas the corresponding expectation value of $\hat{S}^2_{tot}$ is reported in Fig.~2(c). The expectation value of $\hat{S}^2_{tot}$ on the ground and first excited quasi-energy states is computed as a time-average over one oscillation cycle. However, in the high-frequency regime of our simulations, the periodic part of the quasi-energy state (i.e. the Floquet mode) is composed of a dominating dc term plus an ac correction made of small harmonic terms (see e.g. \cite{Grifoni_PR_98} and references therein). Therefore, the total spin of the system in the quasi-energy states is almost stationary. Note that as the interaction energy approaches a critical value $U_C \simeq 0.43 t_1$, the first two quasi-energies attain a level crossing, and the ground quasi-energy state experiences a transition from anti-ferromagnetic (AFM) ordering, with $S_{tot} = 0$, to a saturated ferromagnet (FM), with $S_{tot}=1$.
The accuracy of the asymptotic analysis can be checked by comparing the behaviour of the driven lattice with the equivalent static (undriven) lattice with renormalized tunneling rates $t'_1$ and $t'_2$ given by Eqs.~(12) and (13) with $\Gamma = 2.2$. Numerically computed energy spectrum aside with ground-state and first excited state energies and total spin expectation value for the equivalent static lattice are reported in Fig.~3. The excellent agreement with the results of Fig.~2 confirms the accuracy of our effective Hubbard model with renormalized hopping rates.
The above results suggest the possibility to manipulate magnetic ordering in non magnetic systems of strongly-interacting electrons hopping on a one-dimensional tight-binding lattice. Interestingly, a weak low-frequency modulation of the electric field amplitude can be exploited to switch the system between AFM and FM states, once the high-frequency field has driven the system close to the phase transition point. It is worth noting that since $r(\Gamma)$ is not bound from below, our approach offers enough degrees of freedom to attain AFM/FM transition in a broad region of $U$-$t_{2}$ parameters, by acting on the field amplitude. This is illustrated in Fig.~4 for the simple case of $N=2$ electrons discussed above. The figure shows the critical value of the normalized driving field amplitude $\Gamma$ in the plane ($t_2/t_1$,$U/t_1$) at which AFM/FM transition occurs.
\begin{figure}[h]
\begin{center}
\onefigure[width=8cm]{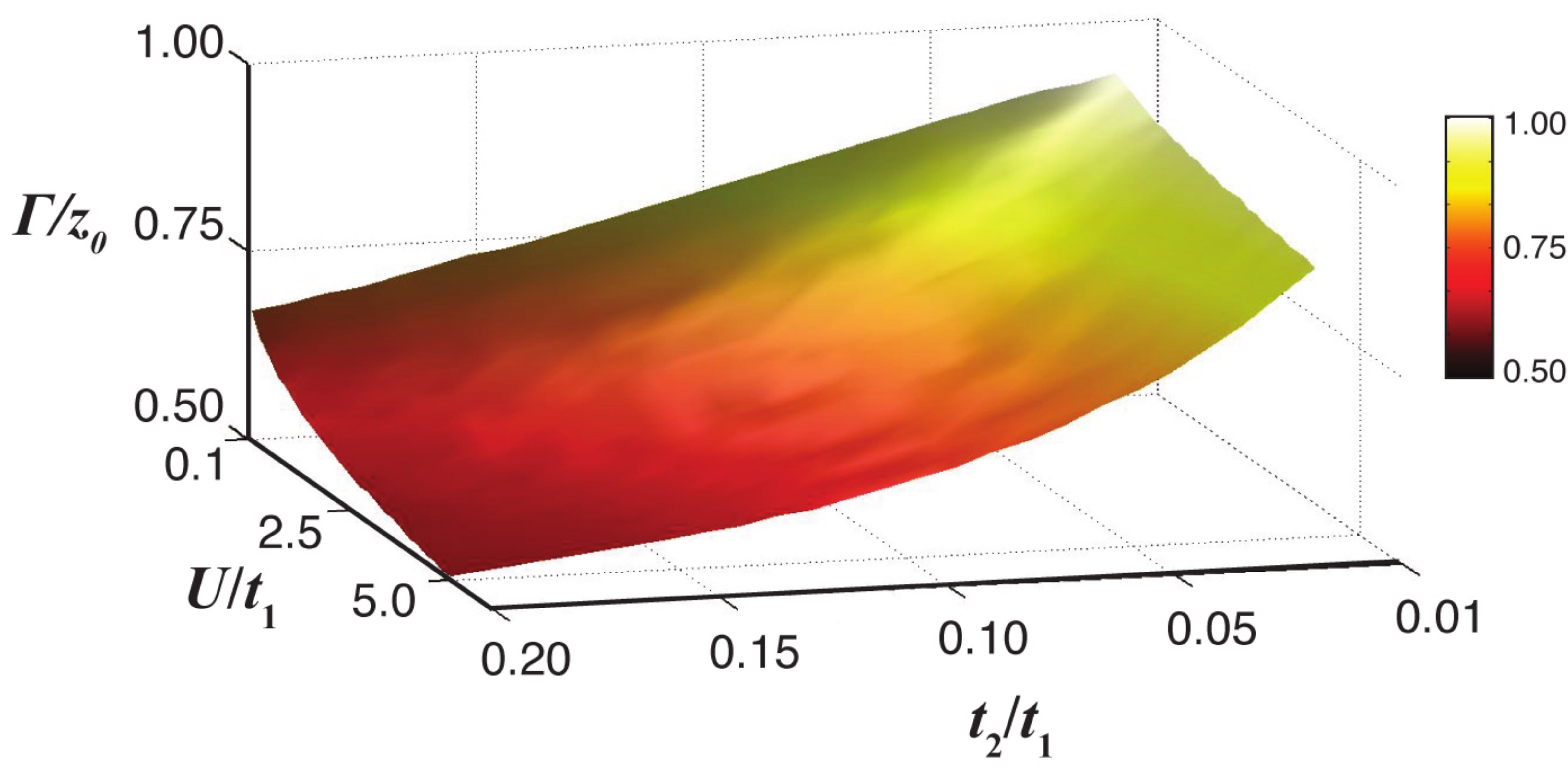}
\caption{Critical value of $\Gamma$, separating AFM and FM phases, as a function of second neighbor tunneling rate $t_2$ and interaction energy $U$, normalized to $t_1$.}
\end{center}
\label{Fig4}
\end{figure}

\section{Conclusions}

In this work we analytically demonstrated that under high-frequency strong electric fields a non-magnetic system of $N$ interacting electrons hopping on a one-dimensional tight-binding lattice with non-zero next-nearest neighbors hopping rate can be driven into a ferromagnetic quasi-energy state. A numerical validation provided for the simple case of $N=2$ electrons with opposite spin on a finite lattice with 11 sites, confirmed the analytical prediction. Our results suggest that the system can be driven to the anti-ferromagnetic/ferromagnetic transition point in a controllable way and for a broad range of parameters. The possibility to actively induce ferromagnetism in non magnetic media can disclose novel opportunities and stimulate further developments in artificial magnetism, with potential applications to spintronics and quantum magnetic devices. Also, the capability of controlling the ratio between first and second neighbor tunneling rates in tight-binding lattices can be profitably exploited in cold atom systems to access quantum simulation of $t_1$-$t_2$ Hubbard ferromagnetism~\cite{Zaleski_JPB_10}.

\end{document}